\documentclass[twocolumn,showpacs,preprintnumbers,amsmath,amssymb,showkeys]{revtex4}

\usepackage{graphicx}
\usepackage{dcolumn}
\usepackage{bm}

\begin{document}

\title{Software graphs and programmer awareness}

\author{G. J. Baxter}
\email{gareth.baxter@mcs.vuw.ac.nz}
\author{M. R. Frean}
\affiliation{
School of Mathematics, Statistics and Computer Science,
Victoria University of Wellington,
P.O. Box 600 Wellington 6140,
New Zealand
}


\date{\today}

\begin{abstract}
Dependencies between types in object-oriented software can be viewed
as directed graphs, with types as nodes and dependencies as edges. The
in-degree and out-degree distributions of such graphs have quite
different forms, with the former resembling a power-law distribution
and the latter an exponential distribution. This effect appears to be
independent of application or type relationship.
A simple generative model is proposed to explore the
proposition that the difference arises because the
programmer is aware of the out-degree of a type
but not of its in-degree. The model reproduces the two distributions,
and compares reasonably well to those observed in 14
different type relationships across 12 different Java applications.
\end{abstract}

\pacs{89.75.Hc, 89.20.Ff, 05.40.-a ,89.75.Fb}
\keywords{networks, object-oriented programming, stochastic processes}
\maketitle

\section{Introduction}\label{intro}\label{data}

Modern computer programs are large and highly structured entities. Naturally
the components of a program depend on each other, and in general if we
think of the
program components as nodes and dependencies as edges, a directed graph can be
constructed.
There is not a single graph for each program, as there are many different
ways that `node' and `edge' might be defined.
For example, for a given {\emph object-oriented} program, we
might construct the graph in which nodes are top-level
types and an edge from type $a$ to type $b$ indicates that type $a$
has a field of type $b$. Thus the number of types of fields declared
in $a$ can be considered as the `out' degree of $a$, while the number of types
having fields of type $a$ is its `in' degree. 
As another example, two top-level types could be linked
when one contains a method (out) with a parameter of the other's type
(in). These different ways of constructing the graph will be referred to
as different \emph{metrics}.

The distributions of in-degree and out-degree show a clear difference
in form. This dimorphism was observed in a
range of graphs generated from the source code of a large corpus of
open-source Java software in \cite{baxter06}. It was found that the
in-degree distributions were well fitted by power-law distributions,
which appear as a straight line when plotted on logarithmic axes.
The
out-degree distributions, on the other hand, are noticeably curved on
a log-log plot.

 This pattern
appears regardless of the metric used or the application of the
software examined
\cite{baxter06,yan06}, and even appears in
other kinds of software-derived graph structures
\cite{myers03,potanin05,valverde05}.
Software code is a direct product of the actions of
programmers, and therefore the resulting `shapes' of the code result
from these actions. It is clear that the difference between in-degree
and out-degree distributions is not accidental,
but is in some way a result of the way in which nodes and edges
are created as the program is written. Because this basic shape is
observed in such a variety of software and metrics, the mechanism must
be quite general, and cannot depend on any specific features of the
way different kinds of dependencies are created or different design
methodologies are used to write software.

While we may be able to characterize the different distributions by
fitting functions of different kinds, and examining the parameters of
the fitted functions for trends and patterns -- for example, the
exponent of a power law distribution -- such descriptive models
 can never explain what we see at any deep level. We
really want to know not just what shapes software has, but how these
shapes come about.
One explanation, suggested in \cite{baxter06}, is that the out-degree
is more actively controlled by the programmer. The outgoing edges of a
type consist of references directly coded as the type is written, while for the
in-degree, references to a type are created as \emph{other} types are
written.

In this paper, we propose a simple
generative model based on this observation which 
reproduces features observed in real software
graph degree distributions. The model aims to capture the simplest actions of a
programmer with respect to type and dependency creation. New edges are
added between existing nodes of the graph, and a new node can be
created by the division of an existing node into two parts. The
treatment of incoming and outgoing connections between nodes is
symmetrical in every way except that the division of nodes depends on
the out-degree of the parent node in a specific way, but is
independent of the node's in-degree. This represents the
programmer's awareness of the outgoing dependencies but not of the
code elsewhere in the
program which refers to the current type. It is found that this single
asymmetry is sufficient to reproduce the difference in shape between
the two degree distributions observed
in real software.
The model proposed is similar to that developed by Price
  \cite{price76} to explain citation rates for papers, though whereas
  Price's model produces power-law
  distributions for both in-degree and out-degree distributions, the
  introduction of the splitting step converts the
  out-degree distribution to an exponential distribution.


\begin{figure*}[htbp]
\includegraphics[width=0.3\textwidth]{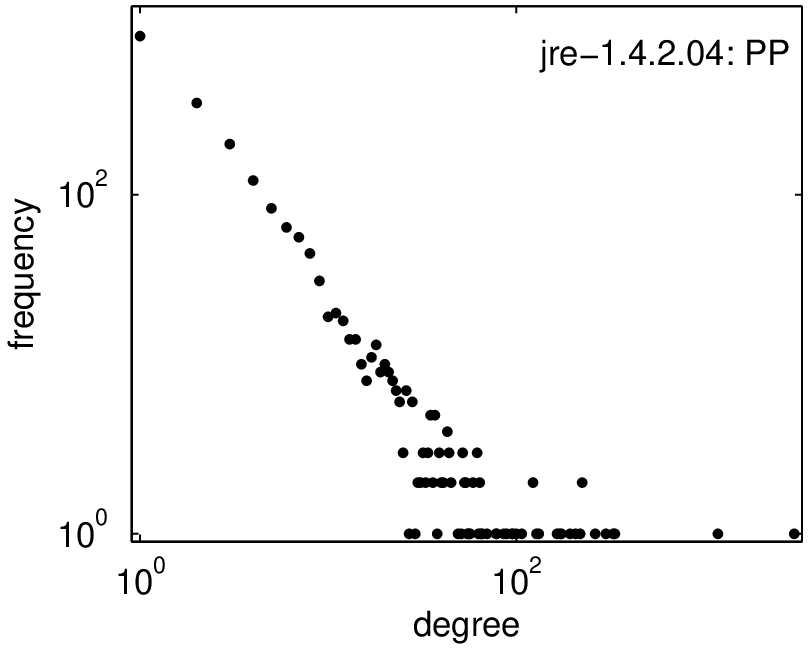}
\hfill
\includegraphics[width=0.3\textwidth]{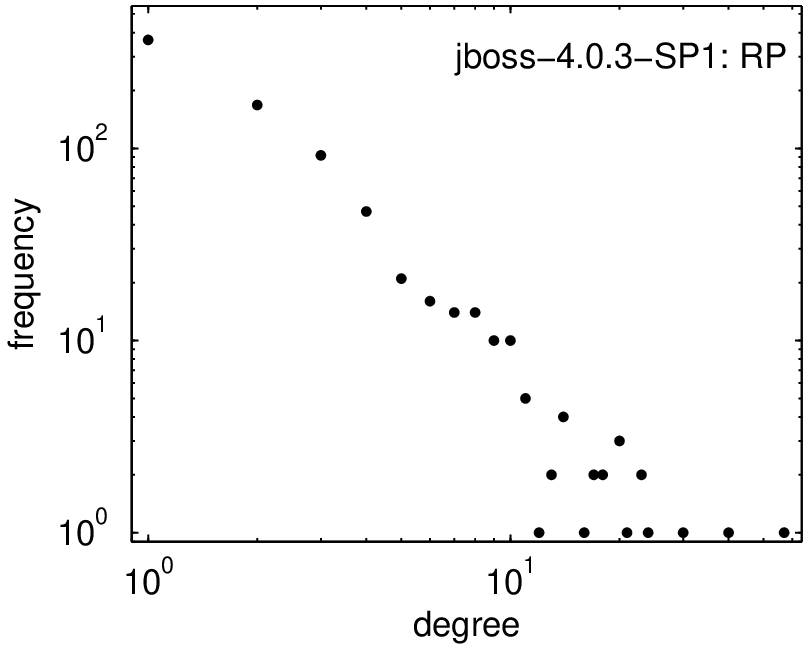}
\hfill
\includegraphics[width=0.3\textwidth]{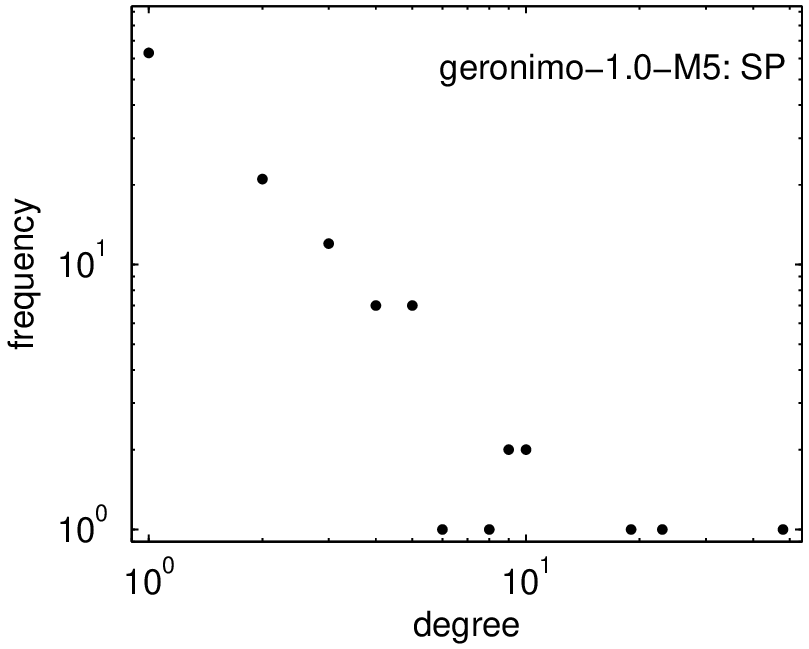}
\caption{Some examples of degree distributions for `in' metrics that
  appear to have a power-law distribution. Both axes are
  logarithmic. Power law distributions appear straight on these
  axes.}\label{FigINsamples}
\includegraphics[width=0.3\textwidth]{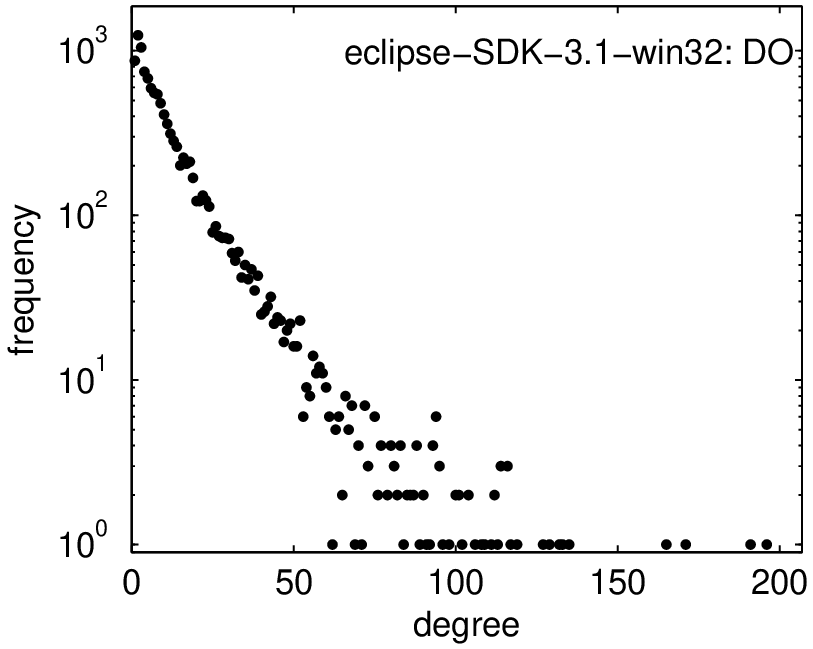}
\hfill
\includegraphics[width=0.3\textwidth]{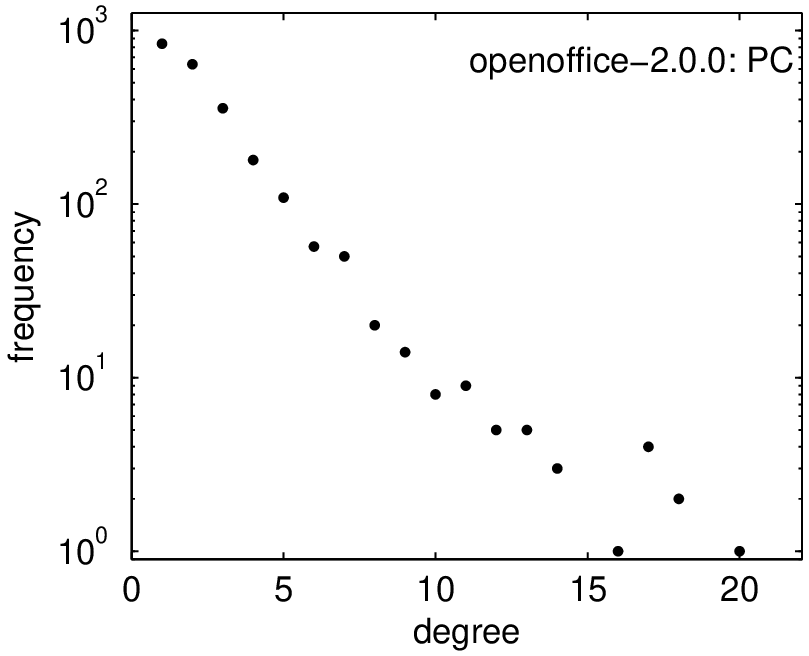}
\hfill
\includegraphics[width=0.3\textwidth]{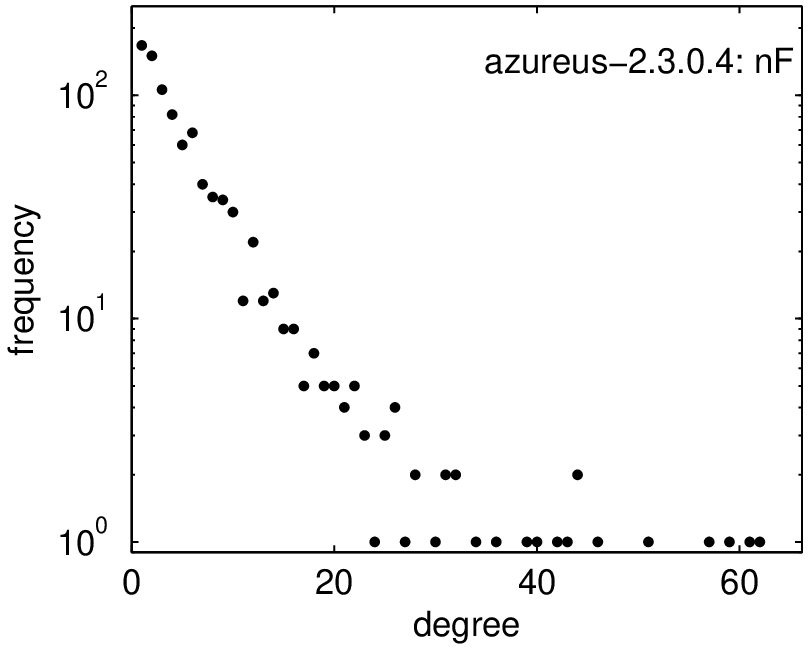}
\caption{Some examples of degree distributions for `out' metrics that
  appear to have an exponential distribution. $x$-axis is linear,
  $y$-axis is logarithmic. Exponential distributions appear straight
  on these axes.}\label{FigOUTsamples}
\end{figure*}

We consider the degree distributions of a variety of inter-class
relationships in the Java source code of 12 different applications.
 These 12 programs
  are the largest of the 50 studied in \cite{baxter06}.
The actual programs (and version numbers) used are listed in
Appendix~\ref{programs}. 
Each metric counts for
each type 
a certain kind of dependency on other types.
We consider 14 metrics that can be identified
as measuring either an out-degree or in-degree.
 Metrics used in \cite{baxter06} which could not be
classified unambiguously as in- or out-degree distributions were left
out.  The metrics are defined in Appendix~\ref{metrics},
and each is referred to by a short abbreviation.
Some pairs of metrics register opposite ends of the same relationship,
suggesting that they are the out- and in- degree distributions of the
same graph.
However, considerations such as the
separation of application code from shared libraries mean that the
count of outgoing edges does not always match the total of incoming
edges.
We will not go into
detail about the differences between the metrics, as our aim is simply to
demonstrate the extent to which the model described below reproduces
the patterns observed in these various data.

Plots of the 5 `in' metrics (see Appendix~\ref{metrics}) on logarithmic
axes often have a linear
form, suggesting a power-law distribution is typical for graph in-degree
distributions, regardless of the specific metric used or the
particular program. 
Three examples are shown in Fig.~\ref{FigINsamples}, which shows the
degree distributions for three
different `in' metrics (see Appendix~\ref{metrics}) in three different
programs of differing sizes.
The 9 `out' metrics (see Appendix~\ref{metrics}) do not appear linear
on doubly logarithmic axes,
however a number of the plots do have a linear shape when plotted on
linear-logarithmic axes, suggesting an exponential distribution.

Not all the data sets conform clearly to this pattern, having a
slightly different shape or one or more points which do not fall on a
neat curve.
Nevertheless,
the general pattern for in-degree is power-law like, and for
out-degree seems to be an exponential shape. This
suggests there is a
  common underlying process, modified to a greater or lesser degree by
  specific
  programmer actions in each case. 

Software graph structures have been examined in several recent studies
\cite{myers03, baxter06, wheeldon03, potanin05, valverde05,
  valverde05b, yan06, concas07}. In
particular, several have reported power-law like degree
 distributions in graphs derived from source code
 \cite{myers03,wheeldon03} or from
 object relationships at run-time \cite{potanin05}.
A distinction between in-degree and out-degree distributions has
been observed in graphs derived from C and C++ software by
Myers \cite{myers03}, who treated both as approximate
power-law distributions, and Valverde and Sol\'{e}
\cite{valverde05},who in common with the present study of Java
software, characterized
the in-degree as a power-law and the
out-degree as an exponential distribution. They showed that these
distributions can be generated by certain cases of the GNC (`growing
network model with
copying') network growth model \cite{krapivsky05}, although the power
law distribution generated by this model has a fixed exponent of 2.
Yan, Qi and Gu \cite{yan06} examined  Java
applications, constructing the directed graph of `import'
relationships. Once again they note that the in-degree\footnote{Our
  convention. Yan et al. use the opposite convention for the
  direction of edges in the graph.} distribution typically resembles a
power-law while the out-degree has a largely exponential
behavior. Concas et al. \cite{concas07} also studied a Java
application, and noted a difference between in-degree and out-degree
distributions.
These observations are confirmed by the present study of a much
larger group of Java applications and metrics.
Yan et al. also postulate a generative model for such
distributions, but make no claims about its plausibility in relation
to programmer actions.

\begin{figure*}[htp]
\includegraphics[width=0.28\textwidth]{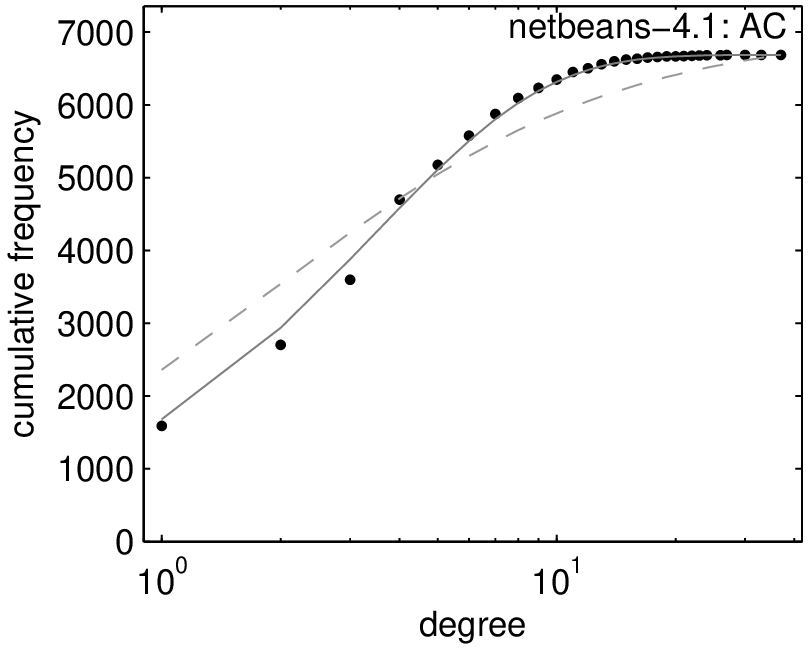}
\hfill
\includegraphics[width=0.28\textwidth]{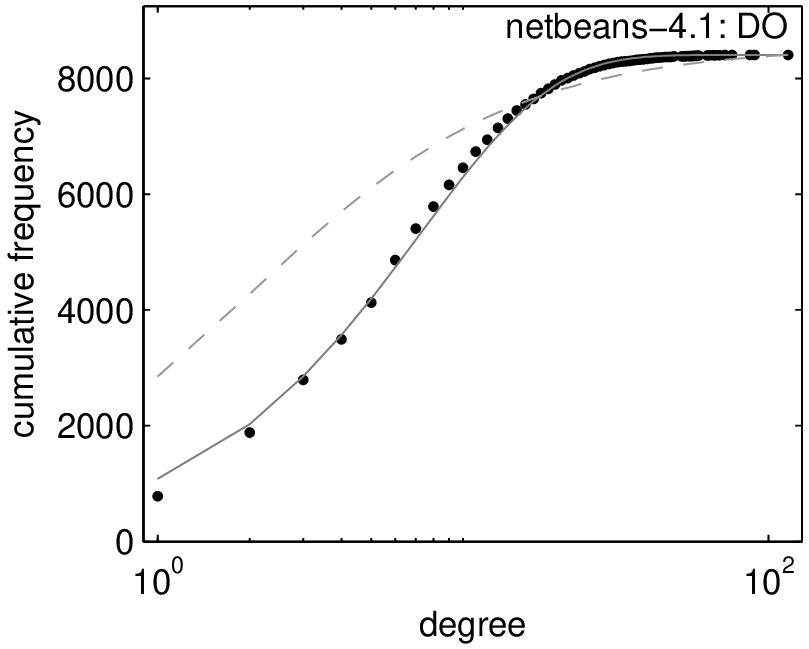}
\hfill
\includegraphics[width=0.28\textwidth]{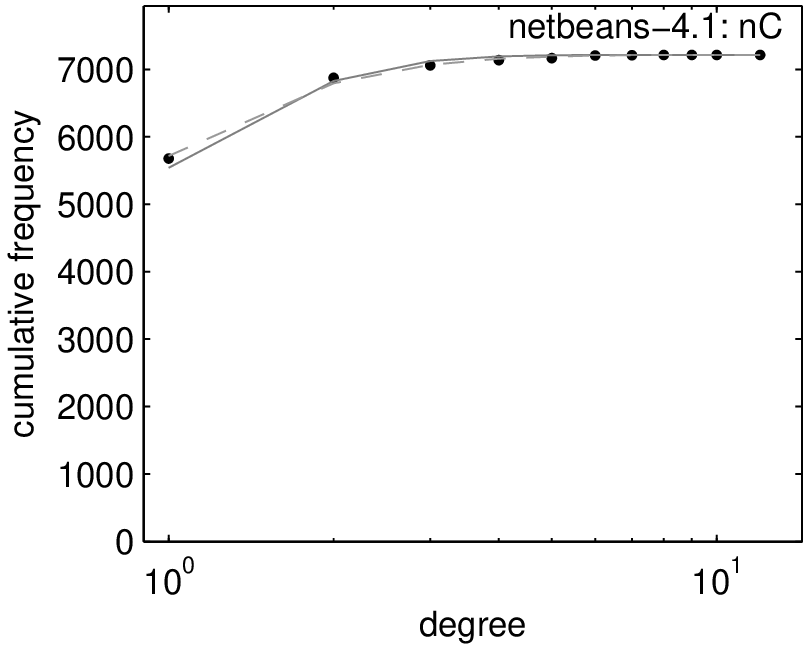}
\\
\includegraphics[width=0.28\textwidth]{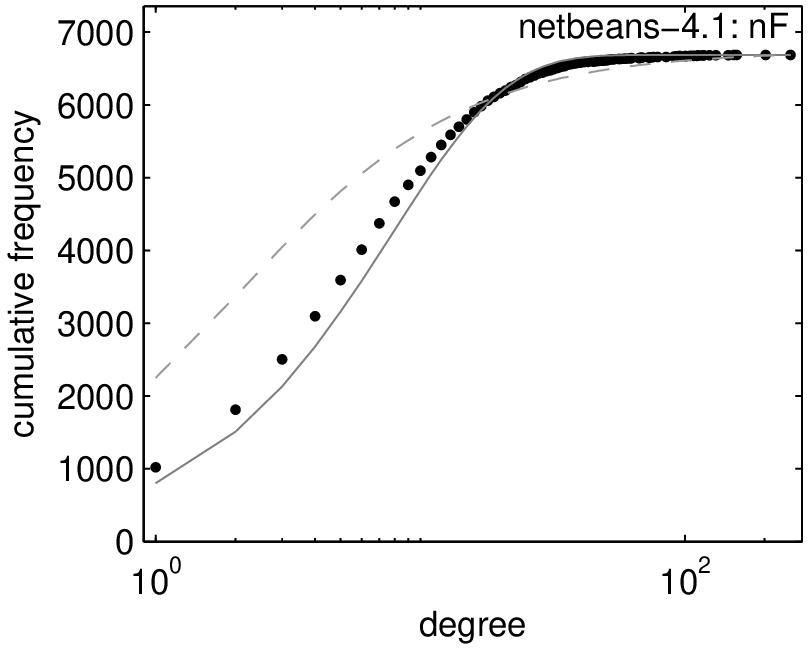}
\hfill
\includegraphics[width=0.28\textwidth]{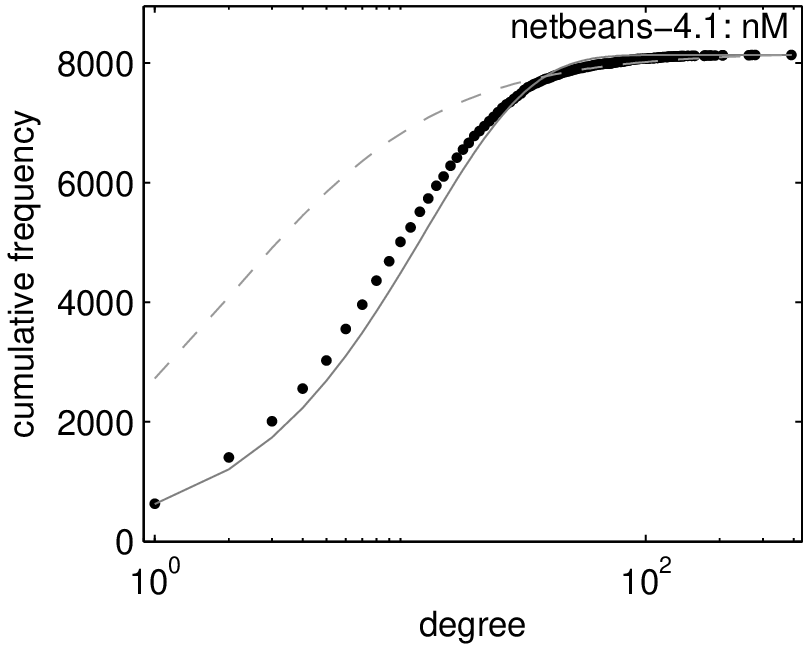}
\hfill
\includegraphics[width=0.28\textwidth]{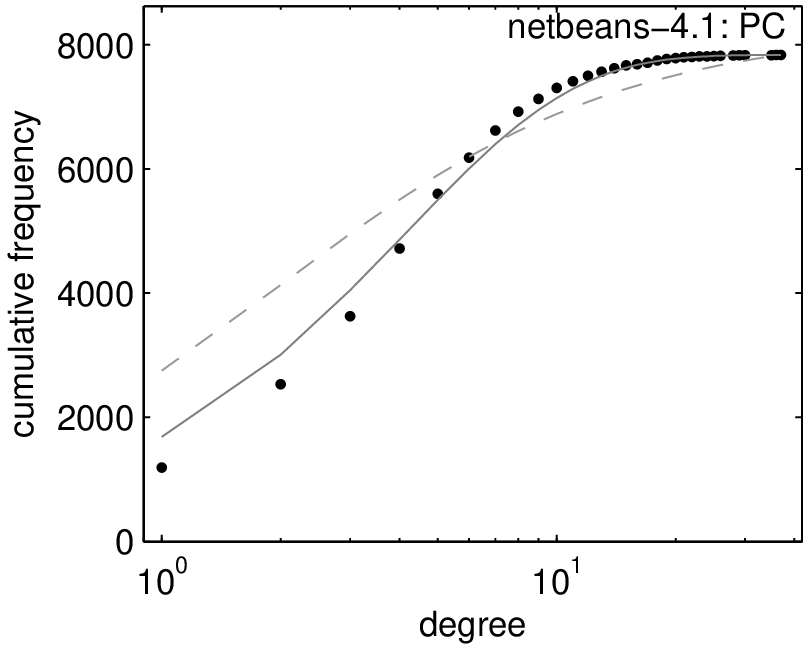}
\\
\includegraphics[width=0.28\textwidth]{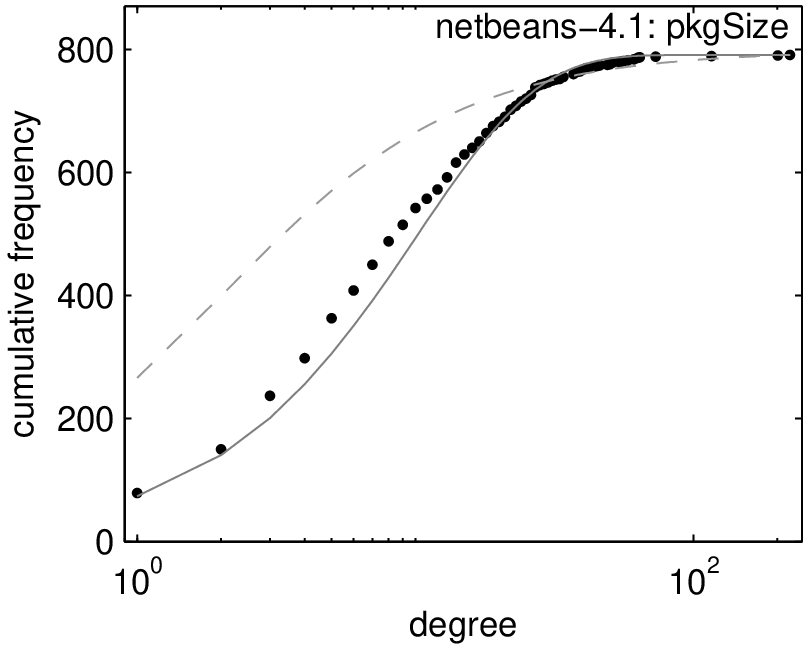}
\hfill
\includegraphics[width=0.28\textwidth]{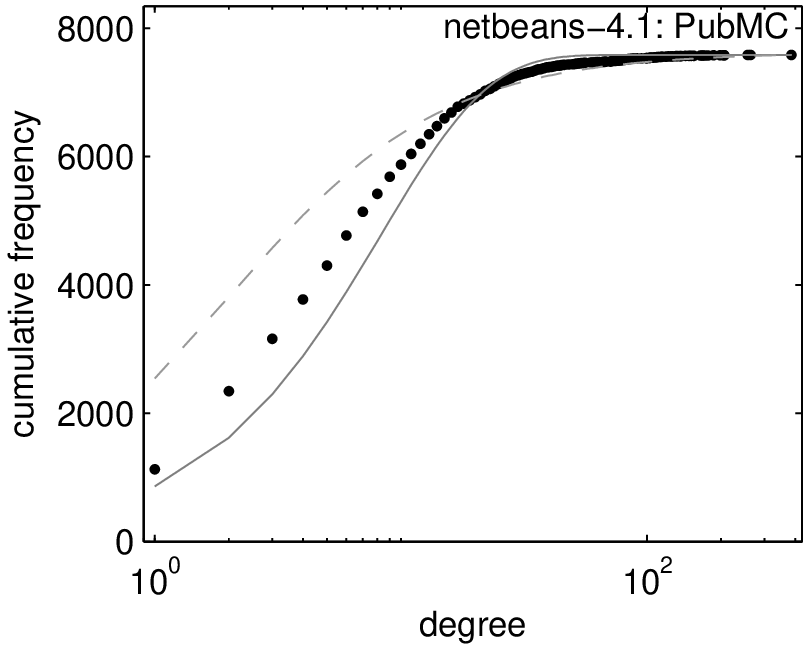}
\hfill
\includegraphics[width=0.28\textwidth]{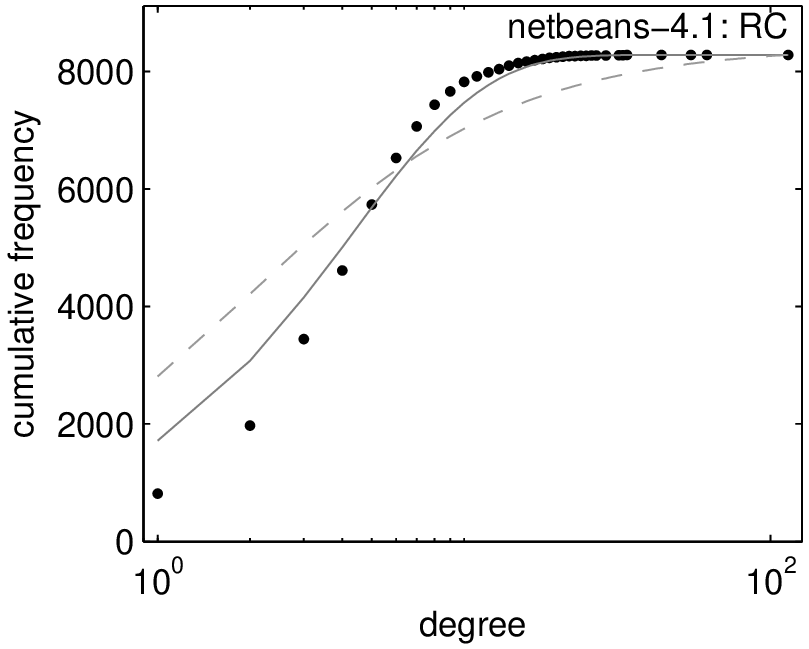}
\caption{Cumulative degree distribution of `out' metrics (as labeled, see
  Appendix for definitions) for
 {\tt netbeans-4.1} with best fit model out-degree
 distribution (solid gray line) as described in Section
 {\ref{model}}. Also shown for comparison is the best fit model
 in-degree distribution (dashed line).
Horizontal scale is logarithmic and vertical scale is
 linear.}\label{FigOutAll}
\includegraphics[width=0.28\textwidth]{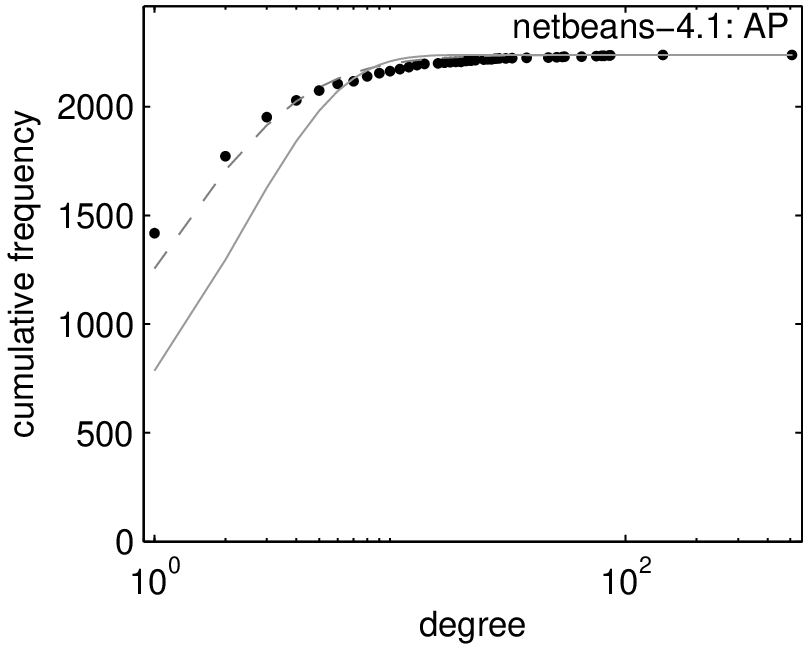}
\hfill
\includegraphics[width=0.28\textwidth]{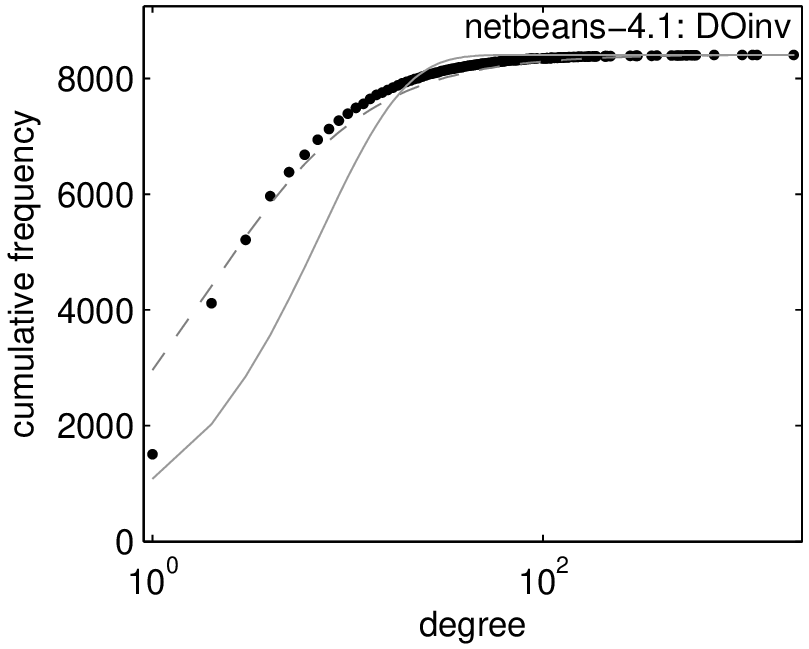}
\hfill
\includegraphics[width=0.28\textwidth]{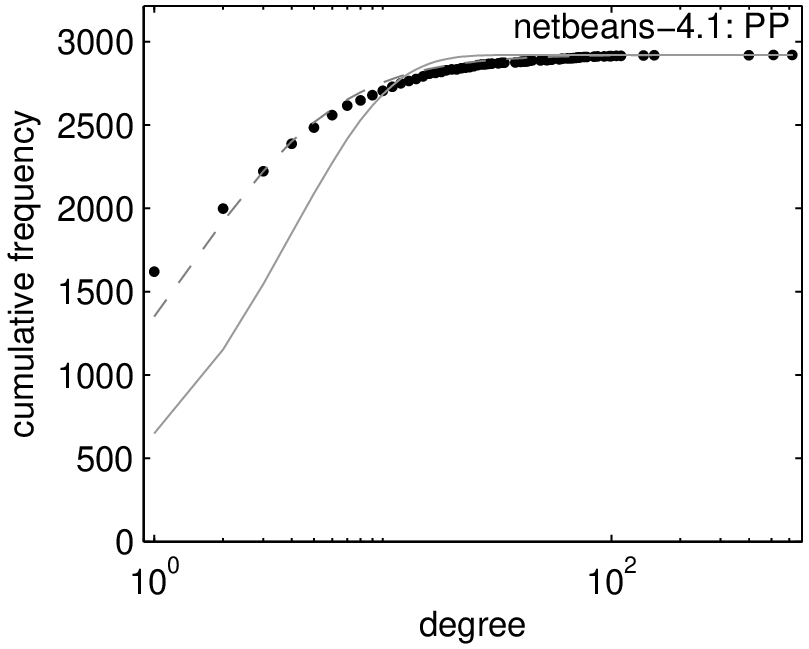}
\\
\includegraphics[width=0.28\textwidth]{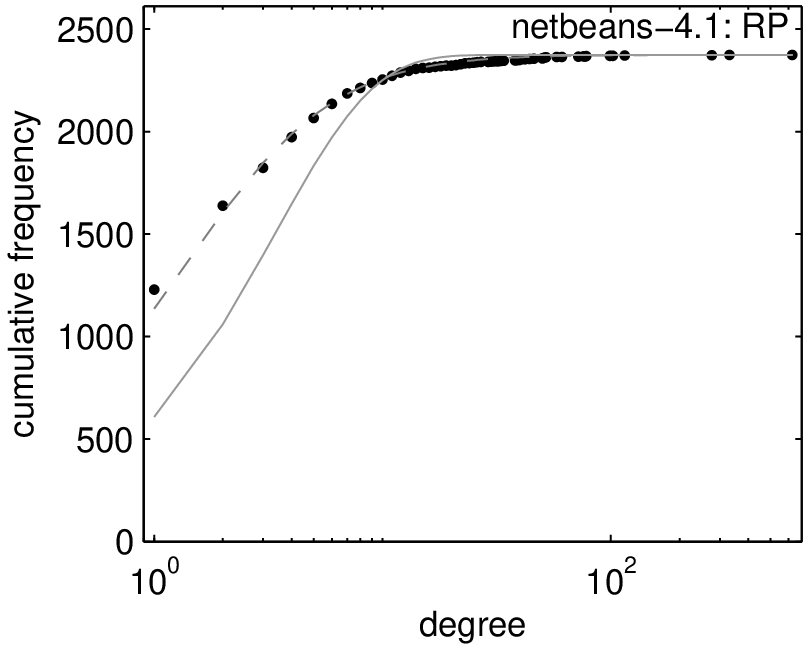}
\hfill
\includegraphics[width=0.28\textwidth]{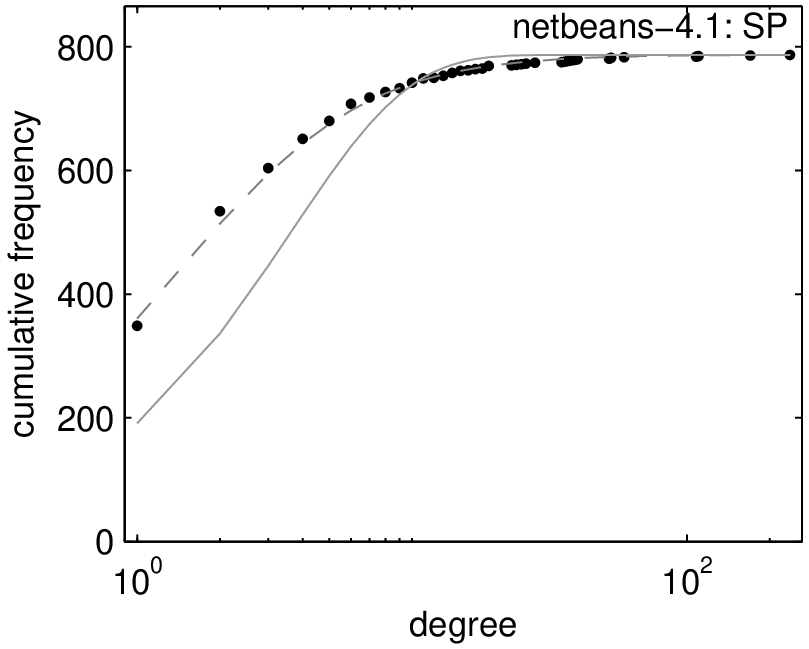}
\hfill
\hspace{0.28\textwidth}
\caption{Cumulative degree distribution of `in' metrics (as labeled, see
  Appendix for definitions) for the largest program studied,
 {\tt netbeans-4.1} with best fit model in-degree
 distribution (dashed gray line) as described in Section
 {\ref{model}}. Also shown for comparison is the best fit model
 out-degree
 distribution (solid line). Horizontal scale is logarithmic and
 vertical scale is linear.}\label{FigInAll}
\end{figure*}

\begin{figure*}[htbp]
\includegraphics[width=0.3\textwidth]{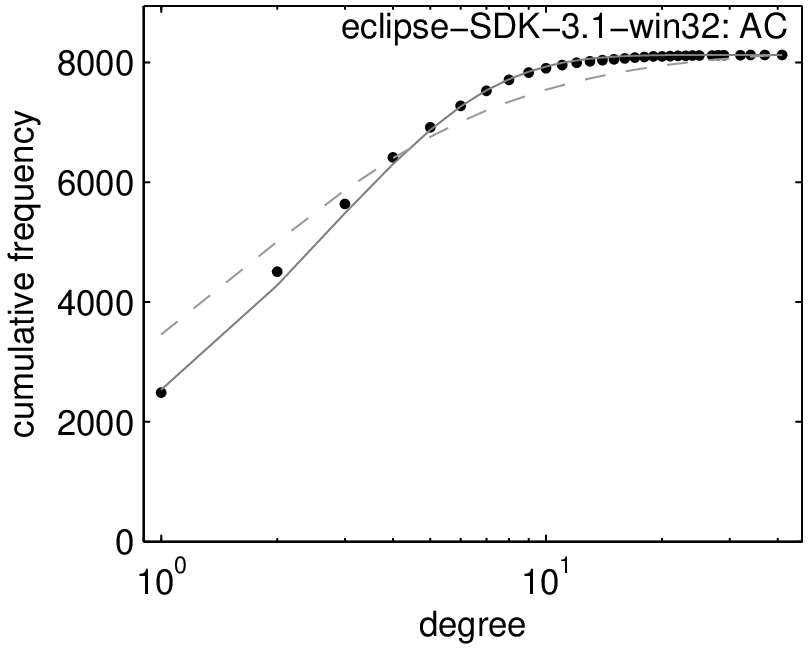}
\hfill
\includegraphics[width=0.3\textwidth]{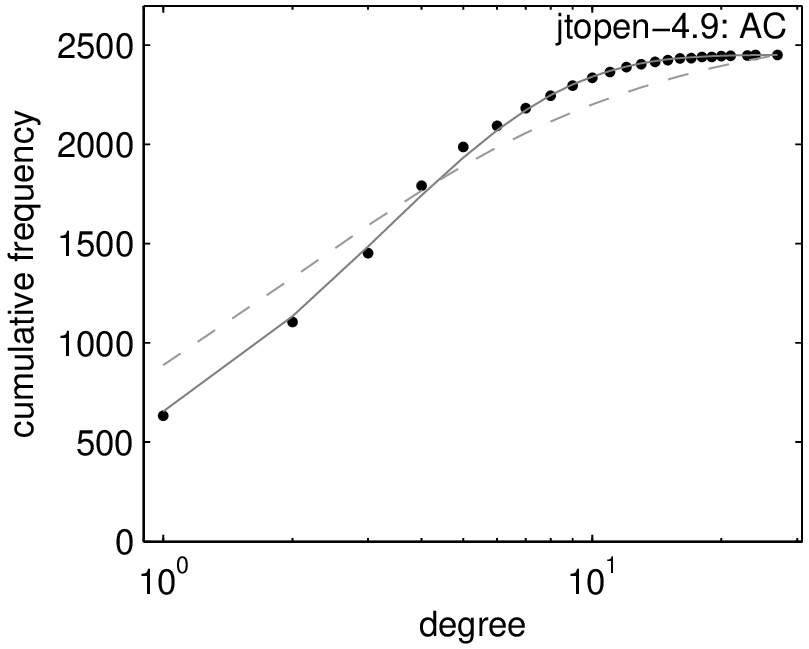}
\hfill
\includegraphics[width=0.3\textwidth]{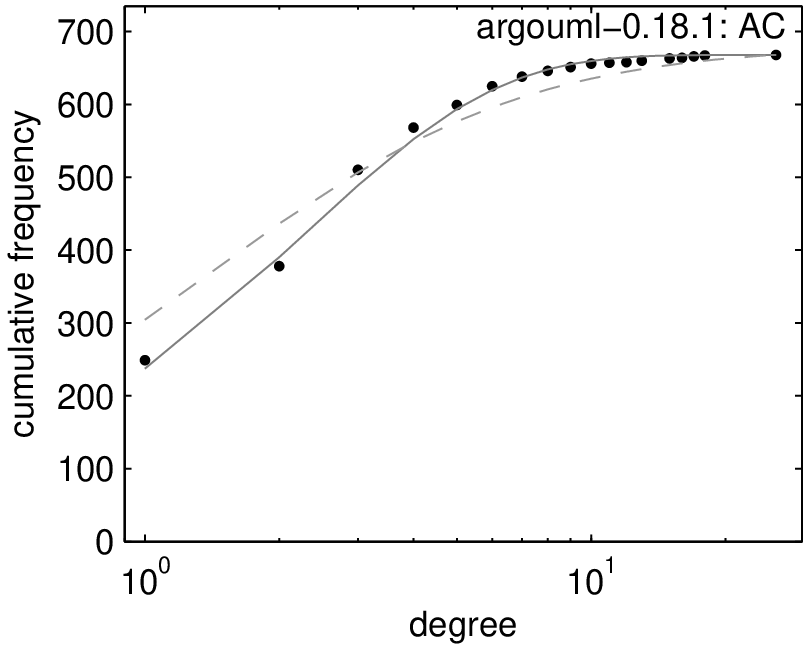}
\\
\includegraphics[width=0.3\textwidth]{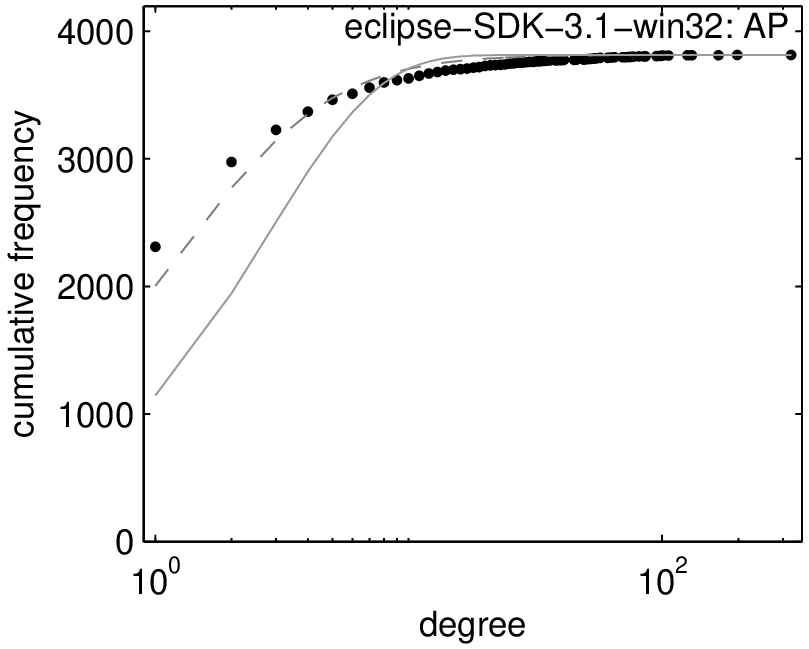}
\hfill
\includegraphics[width=0.3\textwidth]{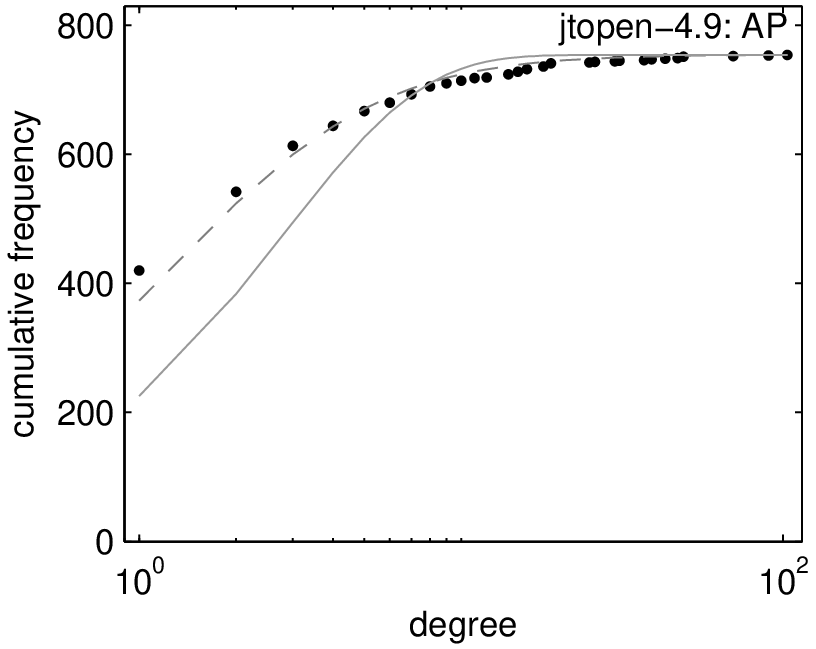}
\hfill
\includegraphics[width=0.3\textwidth]{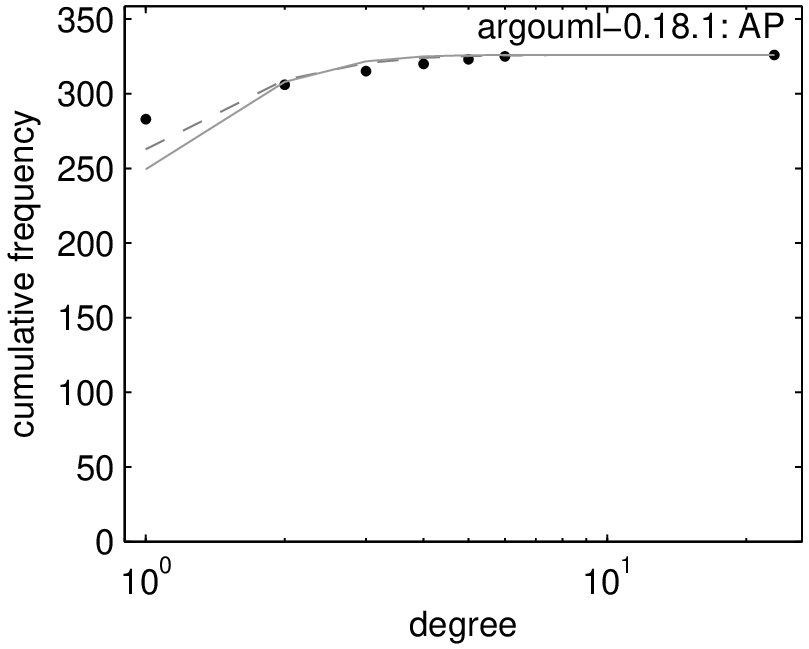}
\caption{Top row: Degree distribution for {\tt AC} (`out' metric) for
  three different programs (as labeled, see Appendix), on
  logarithmic-linear axes.
Bottom row: Degree distribution for {\tt AP} (the reciprocal `in' metric) for
  the same three programs, on logarithmic-linear axes.
}\label{FigACAP}
\end{figure*}

\section{Model}\label{model}

As programmers write software code, they will periodically add
references between types (edges), and occasionally create new types
(nodes). It is these
aspects of the code we are interested in, so the intervening code,
which is actually the majority of the program and specifies all its
functions, is ignored. We consider a simplified process, 
each step of which entails either the addition of an edge
between two existing nodes or the addition of a new node.

Generally there are a
few elements that have  many references to other parts
of the program (these might be the most complex types), while there
are many that reference only a few other types. This divergence can be
approximated by ensuring that the ``the rich get richer'', invoking
the `cumulative advantage' mechanism of Price 
\cite{price76}.
New outgoing edges are added to the type which a programmer is
`currently working on', and we consider that the larger types (those
with most out-edges already) are more likely to be added to in future.
At each step a new edge is added, and the node the link originates from is
chosen with a probability proportional to its current out-degree.

Conversely, there are a
few elements that are referenced repeatedly by many of the other parts
of the program (we might think of these as the simplest and most
universal elements), while there are many that are used only a few
times (these might be more complex elements, at a higher level of
hierarchy, and therefore are less reusable). 
We consider the number of incoming dependencies a
type has to be an approximation to it's usefulness. Therefore
the node to which a new edge will be linked is chosen with a
probability proportional to its current in-degree.

The programmer is conscious of
the number of outgoing references he or she is adding to a node, and at
some point may decide the type is `too big' and create a new one. This
effect is represented by allowing new nodes to occasionally 
be created by `splitting' an existing node into two pieces. The edges
attached to the original node are divided between the two resulting
nodes with each possible division between the two nodes of incoming
and outgoing edges equally likely, with the constraint that at least
one outgoing edge must be transferred to the new type, and one
incoming edge must remain in the original type. Finally, if we
think of the new node as carrying out a subset of the `task' originally
intended for the parent node, the two nodes must be connected, so a
new edge is created from the original type to the new type. This also
ensures that at all times every node has at least one incoming
edge and one outgoing edge.

Let $v_i$ be the out-degree of node $i$, and $w_i$ be its in-degree,
with $i$ running from $1$ up to $k$, the current 
number of nodes. All of these quantities can grow as the process 
proceeds. Let $t$ be the number of steps
carried out so far. Since exactly $1$ edge is added at each step,
$\sum_i v_i = \sum_i w_i =t$.
The process is initiated at $t=1$ with a single node (type) with a single
reference to itself (this is necessary, as links are only added to
nodes which already possess links), i.e. $k=1$ and $v_1=w_1=1$.
At each step:

\begin{itemize}
\item select \emph{parent} node $m$ with probability $v_m/t$
\item  With probability $(1-\gamma)$ simply add an edge:
\begin{itemize}
\item[$\circ$] the parent node $m$ is the \emph{out} node,
\item[$\circ$] select \emph{in} node $n$ with probability $w_n/t$,
\item[$\circ$] $v_m\to v_m+1$ and $w_n\to w_n+1$.
\end{itemize}
\item Otherwise, with probability $\gamma$, split the parent node:
\begin{itemize}
\item[$\circ$] add a new node $k+1$ (the last node number increments
  from $k\to k+1$),
\item[$\circ$] choose $r$ uniformly from $\{1,...,v_m\}$ and $s$
  uniformly from $\{0,...,w_m-1\}$\footnote{The split proportions $r$
    and $s$ are chosen from different ranges to ensure that all nodes
    have at least one in- and one out- link after the new edge
    from $m$ to $k+1$ is added.}, then\\
$v_{k+1}\to r$, $v_m \to v_m-r+1$;\\
 $w_{k+1}\to s+1$ and $w_m \to w_m-s$.
\end{itemize}
\item Increment the counter $t\to t+1$.
\end{itemize}

These steps are repeated for some predetermined number of steps
$t_{final}$. The entire simulation is defined by only two parameters:
the total number of links required (equal to the number of simulation
steps), $t_{final}$, and the splitting probability,
$\gamma$. Since new types only appear due to the
splitting process, $\gamma$ determines the ratio between the number of
types and the number of links: $\frac{k}{t}\to\gamma\mbox{ as }
t\to\infty$.
Note that although nodes to be `worked on' are selected according to
their out-degree, this model is actually symmetric with respect to in-
and out-degree, except for the splitting step: nodes acquire outgoing
edges at a rate proportional to their existing out-degree, and acquire
incoming edges at a rate proportional to their in-degree.
In this model edges are added one by one to different
parts of the graph, so all the types grow at the same time.
New nodes are also created during this process. This doesn't
necessarily reflect the 
actual order in which program code is written.

Although the graph continues to
grow, after a sufficient number of steps, the relative degree
frequencies -- normalized by the total number of nodes --
reach an equilibrium distribution.
Let $C_m$ be the number of
types with out-degree $m$ after step $t$. Considering the two
processes involved in the model, $C_m$ can increase by 1 if an
outgoing edge is added to a type with out-degree $m-1$ and is not
split, or if a type with out-degree
greater than $m$ is split at just the right place that one of the
resulting types has out-degree $m$. Similarly, $C_m$ decreases if a
type of size $m$ gains a new out-link, or is split, so long as the
point of splitting is not $1$ or $m$.
With a little consideration, we can write down the
expected change in $C_m$ at the next step:
\begin{eqnarray}\label{EqDeltaC}
\langle\delta C_m\rangle = (1-\gamma)(m-1)C_{m-1} + \gamma\sum_{r>m}
rC_r\frac{2}{r}\nonumber\\
 - (1-\gamma)mC_m - \gamma mC_m(1-\frac{2}{m})\;.
\end{eqnarray}
The expected fraction of types that have degree $m$ is
\begin{equation}\label{EqDef_f}
f_m=\frac{\langle{C_m(t)}\rangle}{\langle k(t)\rangle}\;.
\end{equation}
It follows from the definition that $\langle\delta C_m\rangle=\gamma
f_m$. Substituting back into
(\ref{EqDeltaC}) we find after collecting like terms that
\begin{equation}\label{EqDeltaF}
(1+m-2\gamma)f_m = (1-\gamma)(m-1)f_{m-1} + 2\gamma\sum_{r=m+1}^R
f_r\;.
\end{equation}
This equation is valid for all $m>1$.
Replacing $m$ by $m-1$ in (\ref{EqDeltaF}), rearranging for the
summation term and substituting back into the original version
of (\ref{EqDeltaF}) gives $f_m$ in
terms of $f_{m-1}$ and $f_{m-2}$
, and after calculating $f_1$ and $f_2$ explicitly we find by
induction the solution for
general $m$ to be
\begin{equation}\label{Eq_f}
f_m=\gamma(1-\gamma)^{m-1}\;,
\end{equation}
which can be written as an exponential
$f_m=\frac{\gamma}{1-\gamma}e^{-\beta m}$ where $\beta=-\ln(1-\gamma)$.

\begin{figure}[htb]
\includegraphics[width=0.38\textwidth]{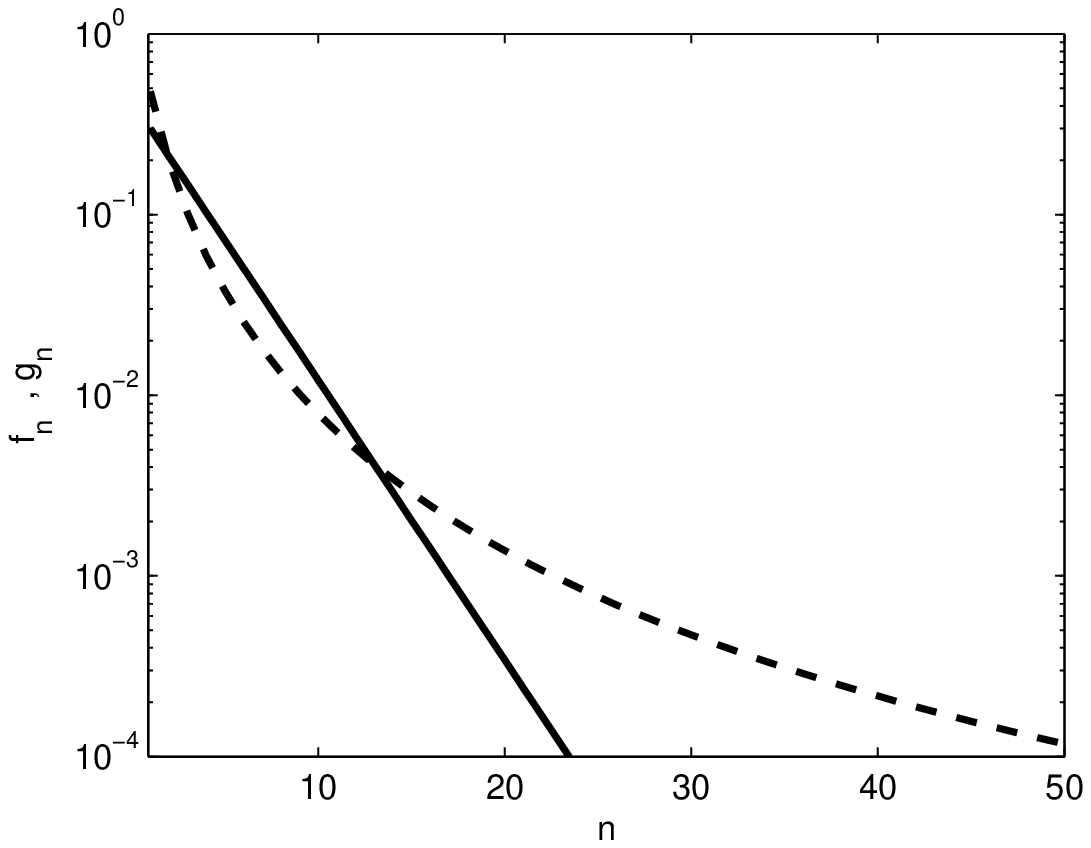}
\includegraphics[width=0.38\textwidth]{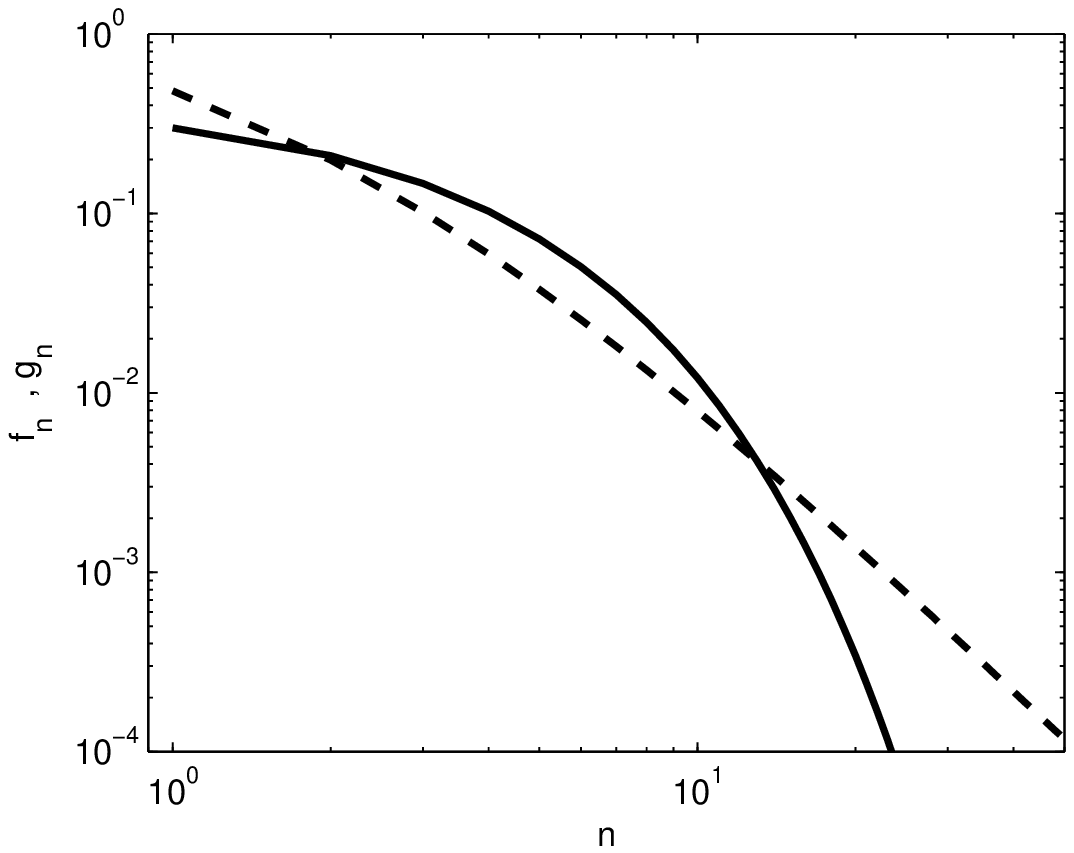}
\caption{Examples of the degree distributions generated by the
  model, with parameter $\gamma=0.3$. Solid line is the out-degree
  model,$f_n$, dashed line the in-degree
model, $g_n$. On linear-logarithmic axes, top, and double
logarithmic axes, bottom.}\label{FigModl}
\end{figure}

A similar calculation can be performed for the in-degree distribution.
The in-degree and out-degree of a type are completely
independent, so the selection of a type for splitting is
uniform with respect to in-degree. If $g_n$, in analogy to $f_m$, is the
fraction of types with in-degree $n$ we find that
\begin{equation}\label{EqDeltaG}
g_n\left[2\frac{n-1}{n}+(1-\gamma)n\right] = 2\sum_{l>n}\frac{g_l}{l} + (1-\gamma)(n-1)g_{n-1}\;.
\end{equation}
Using a similar method to before we find
\begin{equation}
g_n=\frac{(1-\gamma)n}{2+(1-\gamma)n}g_{n-1}\;,
\end{equation}
so that
\begin{equation}\label{Eq_g}
g_n=\frac{\Gamma(n+1)\Gamma(2+\frac{2}{1-\gamma})}{\Gamma(n+1+\frac{2}{1-\gamma})}g_1\;,
\end{equation}
and normalization can be used to find $g_1$.
For large $n$, the ratio $g_n/g_{n-1}$ tends to $1-\frac{2}{(1-\gamma)n}$,
that is, the in-degree distribution tends to a power-law of the form
$g_n=cn^{-\alpha}$ with exponent $\alpha=2/(1-\gamma)$.
Thus the model predicts a decaying exponential for the out-degree
distribution, and an in-degree distribution with a  power-law tail
with exponent greater than or equal to $2$. Examples of
the two distributions (\ref{Eq_f}) and (\ref{Eq_g}) are shown in
Fig.~\ref{FigModl}.

\section{Comparison with measured data}
\label{comparison}

The distributions (\ref{Eq_f}) and (\ref{Eq_g}) were fitted to
the data described in Section \ref{data} using a maximum likelihood
method, which is asymptotically unbiased \cite{barnsdorff94}
. Given some candidate 
distribution $f(x,\gamma)$, the likelihood
that the histogrammed data $h_i$ at values $x_i$ was generated from
this distribution, given the parameter $\gamma$, is
\begin{equation}
p(\mathbf{h}|\gamma) = \prod_{i=1}^k f(x_i,\gamma)^{h_i}
\end{equation}
and we proceed by finding the value of $\gamma$ which maximizes this
quantity.
In the case of out metrics, $f(x_i,\gamma)$ is given by (\ref{Eq_f}),
and the maximum likelihood estimator (MLE) of $\gamma$ is found analytically
to be 
\begin{equation}
\hat{\gamma} = \frac{k}{t}
\end{equation}
as expected,
where $t=\sum_i x_ih_i$ is the number of edges in the graph, and $k=
\sum_ih_i$ is the number of nodes.
For in metrics, we use (\ref{Eq_g}), for which we have not been able
to find a similar solution so the
MLE of $\gamma$ must be found numerically. We again expect $\gamma=k/t$
because this was assumed in the derivation of (\ref{Eq_g}), although in
some cases this is not the best fit value. The explanation of this is
not known, though this very simple model is not expected to explain
every detail of the data.

\begin{figure}[htp]
\includegraphics[width=0.4\textwidth]{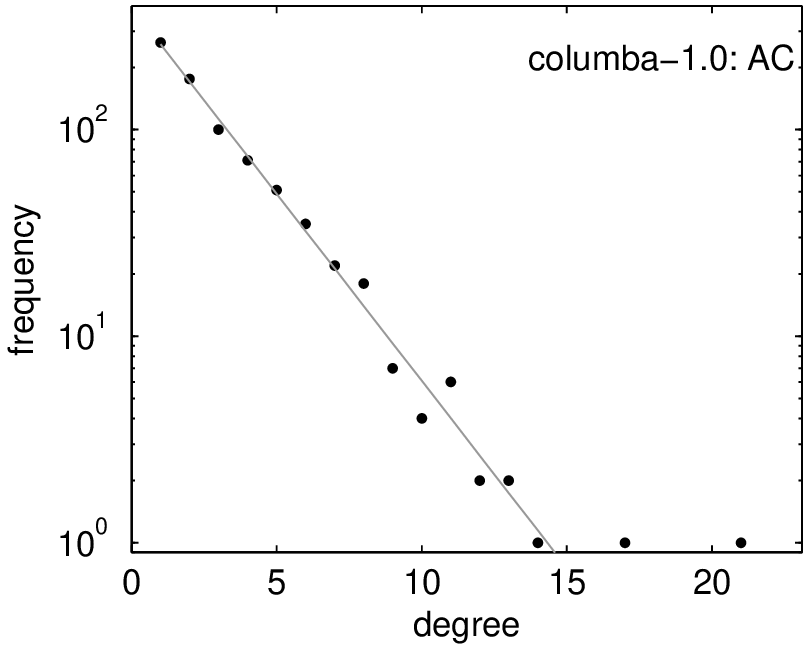}
\caption{{\tt AC} for {\tt columba-1.0}, an example of an out metric
  distribution that is well fitted  by the model.}\label{FigOUTeg1}
\includegraphics[width=0.4\textwidth]{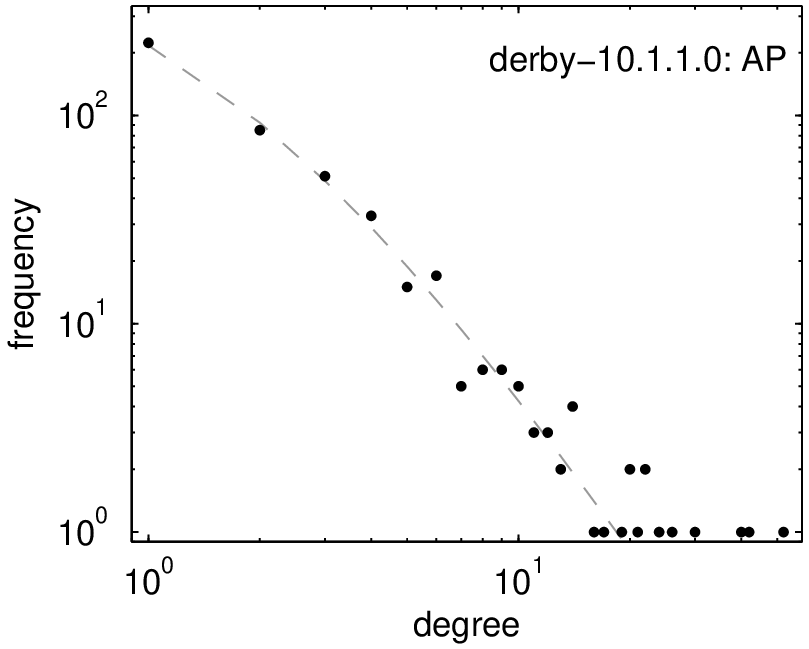}
\caption{{\tt AP} for {\tt derby-10.1.1.0}, an example of an in metric
  distribution that is well fitted  by the model.}\label{FigINeg1}
\end{figure}

The cumulative distribution derived from function (\ref{Eq_f}) using
the best fit (MLE estimated) $\gamma$
value is plotted along  with the out
metric data in Fig.~\ref{FigOutAll}. This
figure plots all of the out-metrics for the same program,
\verb=netbeans-4.1=, and it can be seen that in the majority of cases
the agreement with the data is reasonably good, even though the number
of nodes in the
graphs for different metrics varies widely. Further examples
are shown in the top half of Fig.~\ref{FigACAP}, which shows the same
metric, {\tt AC}, for three different programs. Notice also that a fit of
the `wrong' model distribution (the predicted in-degree distribution)
does not fit as well.
Similarly, the cumulative best-fit functions
(\ref{Eq_g}) for each of the in metrics for
\verb=netbeans-4.1= is plotted 
 in Fig.~\ref{FigInAll}, and for three different programs for the
 same metric ({\tt AP}) in the bottom half of Fig.~\ref{FigACAP}.
 Again,
many data sets show good agreement, and the in-degree model fits the
data much better than the out-degree model (solid curve). 
Comparisons were made
for all the metrics and programs listed in Appendices
 \ref{metrics} and \ref{programs} and Figs. \ref{FigOutAll},
 \ref{FigInAll} and \ref{FigACAP} are fairly representative.
An example of one of the better fits to an out metric data set is
shown in Fig.~\ref{FigOUTeg1}, and an example of a good in metric
fit in Fig.~\ref{FigINeg1}.

To obtain a quantitative measure of how well the model fits the
data, we follow the method of \cite{clauset07} and
calculate a $p$-value (the probability that the data were drawn from
the proposed distribution) based on the Kolmogorov-Smirnov (KS) statistic
\cite{kendall79}
\begin{equation}
D=\mbox{max}|S(x)-F(x)|
\end{equation}
where $S(x)$ is the cumulative distribution function (CDF) of the data and
$F(x)$ is the CDF of the proposed distribution
(i.e. $F(x)=\sum_{y\leq x}f(x)$). The fitted distribution is
correlated with the data, so treating it as the true distribution will
give a falsely high $p$-value. Instead we use a Monte
Carlo procedure, following \cite{clauset07}: A large number of synthetic data
sets is drawn from the best-fit distribution, each one is fitted
individually and the KS statistic (relative to its own best fit
distribution) calculated for each of these fits. The $p$-value is then
the fraction of these KS statistic values that are larger than that
found for the original fit to the real data.

\begin{center}
\begin{table}[htb]
\caption{Number of $p$-values greater than 0.05 and 0.1 for the
out-degree and in-degree data sets.}
\label{tablePvals}
\begin{tabular}{rcccc}
\hline\hline
 & \# $p$-values &  & \# $p$-values  & \\
 &  $> 0.05$ & \% & $> 0.1$ & \% \\
\hline
\multicolumn{1}{l}{\emph{out-metrics}}
 & 4 / 101 & 4\% & 4 / 101 & 4\%\\
\multicolumn{1}{l}{\emph{in-metrics}}
 & 9 / 44 & 20\% & 8 / 44 & 18\% \\
\hline\hline
\end{tabular}
\end{table}
\end{center}

The $p$-values calculated for the in  metrics and out metrics are
tabulated in Table~\ref{tablePvals}. 
We see that quantitatively, the fits are not particularly good. 
 The goodness of
fit for out metrics is particularly poor.
The in metrics are often reasonably well fit by the
model, but where the fit fails, it is sometimes because the degree one point is
lower than predicted -- see for example \verb=Doinv= for
\verb=netbeans-4.1= in Fig.~\ref{FigInAll}. In other
cases, the distribution appears to be
a more pure power law than the model function. Overall, the
heavy tail of the in degree distribution predicted by
the model is successfully observed.
The model fits to out metric distributions are generally worse than
those for the in metrics and often have a more complicated shape,
supporting the hypothesis that the
difference between the distributions is the direct programmer intervention in
the out-degrees of types. 
The KS statistic is the maximum difference between the model and data
cumulative distributions, and we see from Fig.~\ref{FigOutAll} that
although the shapes are similar, there is separation between the two
curves over part of the range in several cases, destroying the
quantitative goodness of fit.
Nevertheless, the model appears to fit many of the curves well or
be qualitatively in agreement for much of the range.

Two assumptions of the model are that types gain edges at a rate
proportional to the existing number of edges, and that when a type is
subdivided the redistribution of
edges between the two resulting types is uniform.
 If the selection of nodes to which new edges are attached is
uniform rather than linear, the resulting in-degree distribution has
an exponential tail, which is not the case in real software graphs, so
it appears that it is necessary that the rate of attachment of new
edges to a node should depend on the existing degree of the node,
though the explanation for this is not as clear as in the case of
wealth distribution or paper citation rates.

In the current model, new types are created by removing references
 (edges) from an
 existing type and placing it in a new type. It is equally plausible
 that a programmer instead copies references, creating duplicates of edges.
 If this is the case, and none of the edges are transferred to the
 new node the out-degree distribution has a `fatter' tail than the
 original exponential distribution, but the `head' of the distribution
 also becomes much steeper, meaning that this model fits the out-degree
 distribution data very poorly. Alternatively if the edges
 are distributed between the two nodes according to a binomial
 distribution (so that each edge is equally likely to be attached to
 either node), the resulting out-degree distribution has a distribution
 similar to an exponential distribution, and the in-degree distribution
 has a power-law tail with a minimum exponent of $3$. This is also
 incompatible with the data, which typically have a
 power-law tail with exponent close to $2$.
 
\section{Discussion}\label{discussion}

In this paper we have described a simple model of the generation of
software graph
degree distributions based on the assumption that the process of
programming involves programmers making
active choices about the structure of the type on which they are
working -- in particular they are conscious of the `size' of the type,
and this comes to be reflected in the resulting out-degree,
while the in-degree of a type emerges indirectly as a result of the
construction of other types. The only difference between incoming and
outgoing edges in the microscopic
process of the model is that the splitting operation on types is
dependent on type out-degrees and is independent of in-degrees. 
This model is extremely simple, and depends on a single parameter
$\gamma$ which can be physically interpreted as the reciprocal of the
average degree of the graph.
This suggests that the mean number of dependencies per type (both
incoming and outgoing) may prove to be a useful statistic for
comparing different type relationships and different programs.
 The model reproduces the approximate
shapes of in-degree and out-degree distributions for a range of graph
construction metrics applied to a variety of Java programs. This
suggests that this shape is due to simple statistical processes common
to all software graphs, so that any differences due to different
design methodologies and so on must be sought in the details of the
deviations from this mean behavior.
The agreement of the proposed model
 with the measured data is not perfect however. These
and other difficulties, such as variations between the shapes of
distributions between programs and between metrics indicates that
there are higher-order effects that remain to be described.

Degree distribution is one of the most accessible measures of the
`shape' of a network, but there are many more measures such as degree
correlation, clustering coefficients and so on  that can be
calculated.
 Further analysis of real data sets for such more detailed
measures would help to discriminate between and to refine
the various generative models that have so far been proposed.
Another avenue of investigation that would be particularly fruitful
would be a statistical analysis of the programming process as it
happens, in order to identify the rates and probabilities of various
actions, which could then be compared with the assumptions of the
model proposed in this paper.

\begin{acknowledgements}
We would like to thank Hayden Melton for compiling the data
used in the analysis and Ewan Tempero and James Noble for
technical advice and helpful discussions.
\end{acknowledgements}

\appendix

\section{metrics}\label{metrics}

Brief descriptions of the metrics used in the study are listed
below. Following \cite{wheeldon03,baxter06} short codes are used
throughout the text to refer to each metric.
For more complete descriptions of the metrics and how the data was
extracted from the source code, see \cite{baxter06}.

Some of the metrics are
paired, with one representing the `in' degree (listed first) and
another representing the reciprocal `out' degree. In general an `out'
metric for a type counts things that would appear in the code for that
type, while `in' metrics count references to a type which appear in
the code for \emph{other} types. Some of the metrics have no
reciprocal metric, but can still be identified as measuring either
`in' or `out' degree.
The remaining metrics used in \cite{baxter06} were not included in
  the analysis as their in/out status was not clear or they were a
  mixture of in and out measures.

\subsection*{`In' metrics}

\begin{description}
\item [{\tt AP} References to class as a member.]
For a given {\bf type}, the number of top-level types (including itself) in
the source that have a field of that type.
\item[{\tt DOinv} Depends On inverse.]
For a given {\bf type}, the number of type implementations in which
it appears in their source.
\item[{\tt PP} References to class as a parameter.]
For a given {\bf type}, the number of top-level types in the source that 
declare a method with a parameter of that type.
\item[{\tt RP} References to class as return type.]
For a given {\bf type}, the number of top-level classes in the source
that declare a method with that type as the return type.
\item[{\tt SP} Subclasses.]
For a given {\bf class}, the number of top-level classes that
specify that class in their {\tt extends} clause.

\end{description}

\subsection*{`Out' metrics}

\begin{description}
\item[{\tt AC} Members of class type.]
For a given {\bf type}, the size of the set of types of fields for that
type.
\item[{\tt DO} Depends on.]
For a given {\bf type}, the number of top-level types in the source that it
needs in order to compile.
\item[{\tt PC} Parameter-type class references.]
For a given {\bf type}, the size of the set of types used as parameters in
methods for that type.
\item[{\tt PubMC} Public Method Count.]
The number of methods in a {\bf type} with public access type.
\item[{\tt RC} Methods returning classes.]
For a given {\bf type}, the size of the set of types used as
return types for methods in that type.
\item[{\tt nC} Number of Constructors.]
For a given {\bf class}, the number of constructors of all access types
declared in the class.
\item[{\tt nF} Number of Fields.]
For a given {\bf type}, the number of fields of all access types
declared in the type. 
\item[{\tt nM} Number of Methods.]
For a given {\bf type}, the number of all methods of all access types (that
is, public, protected, private, package private) declared (that is, not
inherited) in the type.
\item[{\tt  pkgSize} Package Size.]
The number of {\bf types} contained direction in a package (and not
contained in sub-packages).
\end{description}

\section{applications}\label{programs}

The programs studied, their size (number of Classes), domain
and where they were sourced are:

\begin{center}
\begin{tabular}{lclll}
\hline\hline
\multicolumn{1}{c}{\emph{Application}} &
\multicolumn{1}{c}{\emph{\#Classes}} &
\multicolumn{1}{l}{\emph{Domain}} &
\multicolumn{1}{c}{\emph{Origin}} &
\multicolumn{1}{c}{\emph{Notes}}\\
\hline
\mbox{eclipse-SDK-3.1-win32} & 11413 & IDE & {www.eclipse.org} &
Donated by IBM \\
\mbox{netbeans-4.1} 
& 8406 & IDE & netbeans.org & Donated By Sun \\
\mbox{jre-1.4.2.04} & 7257 & JRE & sun.com  & \\
\mbox{jboss-4.0.3-SP1} & 4143 & J2EE server & Sourceforge & \\
\mbox{openoffice-2.0.0} & 2925 & Office suite & openoffice.org  &
Donated By Sun\\
\mbox{jtopen-4.9} & 2857 & Java toolbox for iSeries and AS/400 servers
&
Sourceforge  & Donated by IBM \\
\mbox{geronimo-1.0-M5} & 1719 & J2EE server & Apache & \\
\mbox{azureus-2.3.0.4} & 1650 & P2P filesharing & Sourceforge  & \\
\mbox{derby-10.1.1.0} & 1386 & SQL database & Apache Jakarta  &
Donated by IBM \\
\mbox{compiere-251e} & 1372 & ERP and CRM & Sourceforge  & \\
\mbox{argouml-0.18.1} & 1251 & UML drawing/critic & {tigris.org} & \\
\mbox{columba-1.0} & 1180 & Email client & Sourceforge  & \\
\hline\hline
\end{tabular}
\end{center}


\end{document}